\documentclass{article}  
\usepackage{spadre2008}
\usepackage{graphicx}
\def\etal{\emph{et al.}}
\newcommand{ \sNN }{\sqrt{s_{NN}}}  
\newcommand{ \be }{\begin{equation}}
\newcommand{ \ee }{\end{equation}}
\newcommand{ \bea }{\begin{eqnarray}}
\newcommand{ \eea }{\end{eqnarray}}

\newcommand{ \eps }{\varepsilon}

\newcommand{ \epsp }{\eps_{part}}
\newcommand{ \epspp }{\eps_{PP}}
\newcommand{ \epsrp }{\eps_{RP}}

\newcommand{ \beps}{\mbox{\boldmath$\eps$}}
\newcommand{ \sgmeps }{\sigma_\eps}

\newcommand{ \BG}{\mathrm{BG}}

\newcommand{ \vtwo }{{v_2\{2\}}}
\newcommand{ \vfour }{{v_2\{4\}}}

\newcommand{\mean}[1]{\left< {#1} \right>}

\frompage{000} \topage{000}                                              %|
\def\rightmark{Hints of Incomplete Thermalization}
\def\leftmark{A. Tang et al.}
 
\title{Flow Results and Hints of Incomplete Thermalization} 
\authors{
{Aihong Tang$^1$ for the STAR Colllaboration %
}\\[2.812mm]
{\normalsize
\hspace*{-8pt}$^1$ Brookhaven National Lab,
PO BOX 5000, Upton, NY. USA\\[0.2ex] 
}}
 
\abstract{We classified $v_2$ measurements according to their sensitivities w.r.t. to two planes, namely, reaction plane and participant plane. Likewise, in $v_2/\epsilon$ scaling, we showed that one needs to choose a $\epsilon$ that is sensitive to the same plane as that $v_2$ is sensitive to. We presented our $v_2/\epsilon$ as a function of centrality and transverse momentum. We studied the ratio of $v_4/v_2^2$. We discussed the applicable range for hydrodynamics, as well as implications to an incomplete thermalization. }

\keyword{flow, $v_2$, hydrodynamics, thermalization, RHIC, STAR} 
\PACS{25.75.Ld}
 
\begin{document}
 
\maketitle
\setcounter{page}{1}

\section{Introduction}\label{intro}

In ultra-relativistic heavy ion collisions, spectators pass through each other 
quickly, and the system begins its evolution with what is 
left behind - the overlap region of two nuclei. The pressure gradient convert
the spacial anisotropy, quantified by eccentricity $\epsilon$, of the overlap region into anisotropy in momentum space, quantified by azimuthal anisotropy $v_2$. $v_2$ is defined as the second Fourier coefficient in the description of particles distribution w.r.t. the reaction plane~\cite{methodPaper}, and it is largely determined by the collective motion in the in-plane direction.
$v_2$ at Relativistic Heavy Ion Collider (RHIC) is reported to be large and, for the first time in heavy ion collisions, can be described by ideal hydrodynamics~\cite{STAR1stFlow}. Theoretical calculation shows that in order to explain the large $v_2$ observed at RHIC, one has to assume that the shear viscosity is extremely small~\cite{Teaney}. That is one of the important reasons for which scientists think that a perfect liquid has been formed in relativistic heavy ion collisions~\cite{perfectLiquid}.

Since the announcement of the discovery of a perfect liquid, our understanding of the matter created at RHIC has continued to advance. Elliptic flow analyses have been extended to great details by all four experiments at RHIC. At the same time, due to different techniques used, it becomes an increasingly amount of work to understand/compare results across different experiments and, sometimes, even different analyses within the same experiment. Therefore it becomes important to understand what is the relation between various $v_2$ measurements.
For the physics side, in order to quantify how perfect the liquid is, it is necessary to re-examine the hydrodynamics limit. In this paper, we try to study the hydrodynamic behavior under a more general context, namely, the transport approach which recovers hydrodynamics when mean free path is extremely small if compared to the system size~\cite{OllitraultBoltzmannEquation}

\section{Choosing the right $v_2$ and $\epsilon$ pairs}\label{techno}  

The ratio of $v_2/\epsilon$ reflects how well the initial anisotropy is converted into momentum anisotropy~\cite{v2Scale}. This conversion process is directly affected by dynamics of the system, e.g. Equation Of State, thermalization etc.  It is important to measure this quantity as accurate as possible. However, there exist many $v_2$ measurements, for example,  $v_2$ measured by event plane method ($v_2\mathrm{\{EP\}})$, by cumulants ($\vtwo$,$\vfour$), by Lee-Yang zero method ($v_2\mathrm{\{LYZ\}}$), and by using event plane reconstructed with Shower Maximum Detectors at Zero Degree Calorimeters ($v_2\mathrm{\{ZDCSMD\}}$), etc. Likewise there exist many $\epsilon$ calculations. In this section, we will try to make connections between various $v_2$ and $\epsilon$ methods, and make the justification for the right combination of them.

Define 
\be
  \beps=\{\eps_x,\eps_y\} =
  \left\{ \mean{\frac{\sigma_y^2-\sigma_x^2}{\sigma_x^2+\sigma_y^2}}_{part},
  \mean{\frac{2\sigma_{xy}}{\sigma_x^2+\sigma_y^2}}_{part}  \right\},
\ee
where
$\sigma_{x}^2=\mean{x^2}-\mean{x}^2$, 
$\sigma_{y}^2=\mean{y^2}-\mean{y}^2$, and
$\sigma_{xy}^2=\mean{xy}-\mean{y}\mean{x}$, 
and the average is taken over the coordinates of the participants in
a given event. With this definition, $\eps_x$ is the eccentricity of reaction plane (defined by the impact parameter), and $\eps_x$ is also called $\eps_{RP}$. $\eps_y$ has a distribution centered at zero with finite width. The participant eccentricity measures the asymmetry in the participant plane (defined by the principle axis of the ellipsoid), and is given by  $\epsp=\sqrt{\eps_x^2+\eps_y^2}\equiv \epspp$

\begin{figure}[ht]
\begin{center}
\resizebox{
\textwidth}{!}{
\resizebox*{10cm}{6.6cm}{
\includegraphics{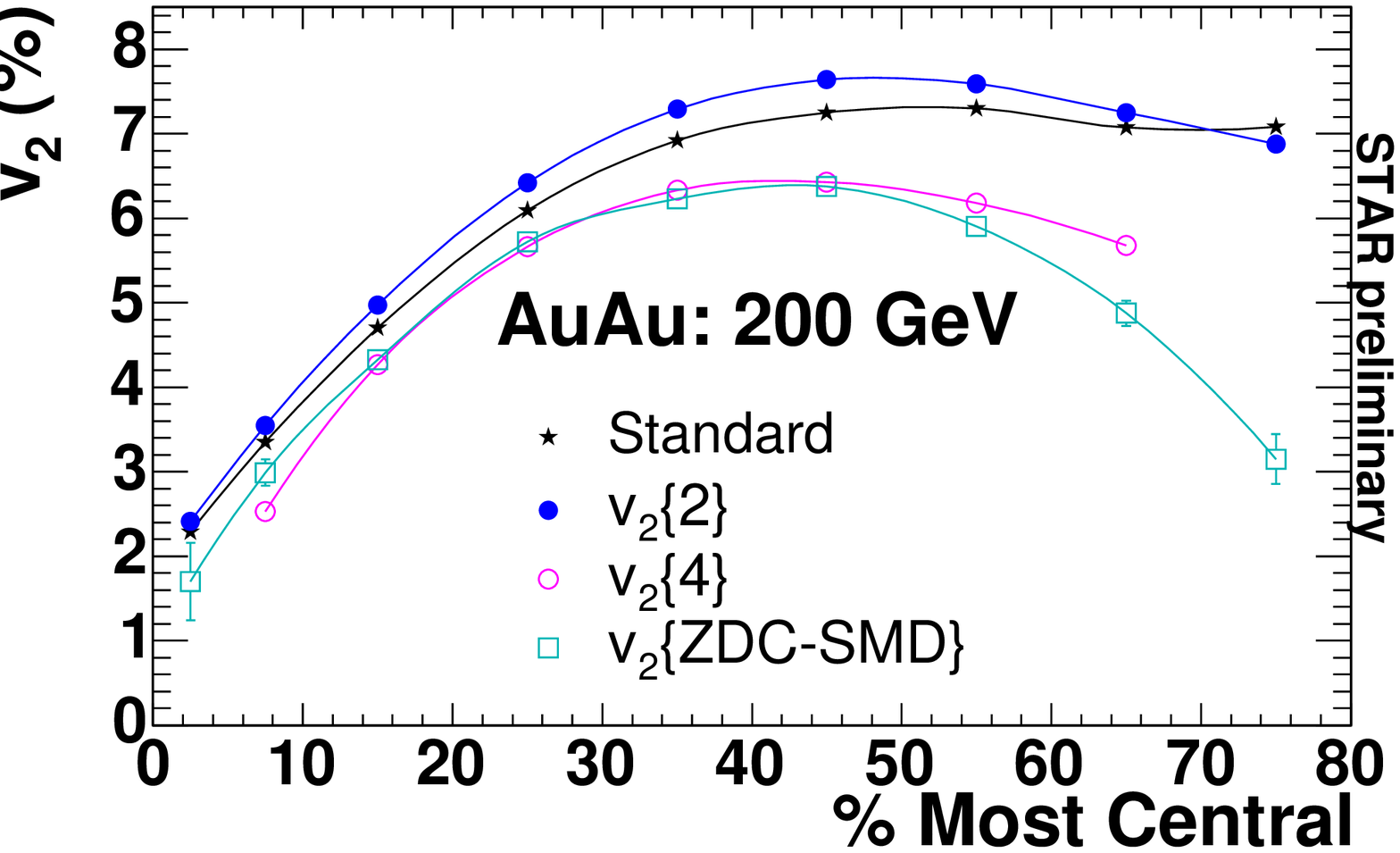}}
\resizebox*{10cm}{7.cm}{
\includegraphics{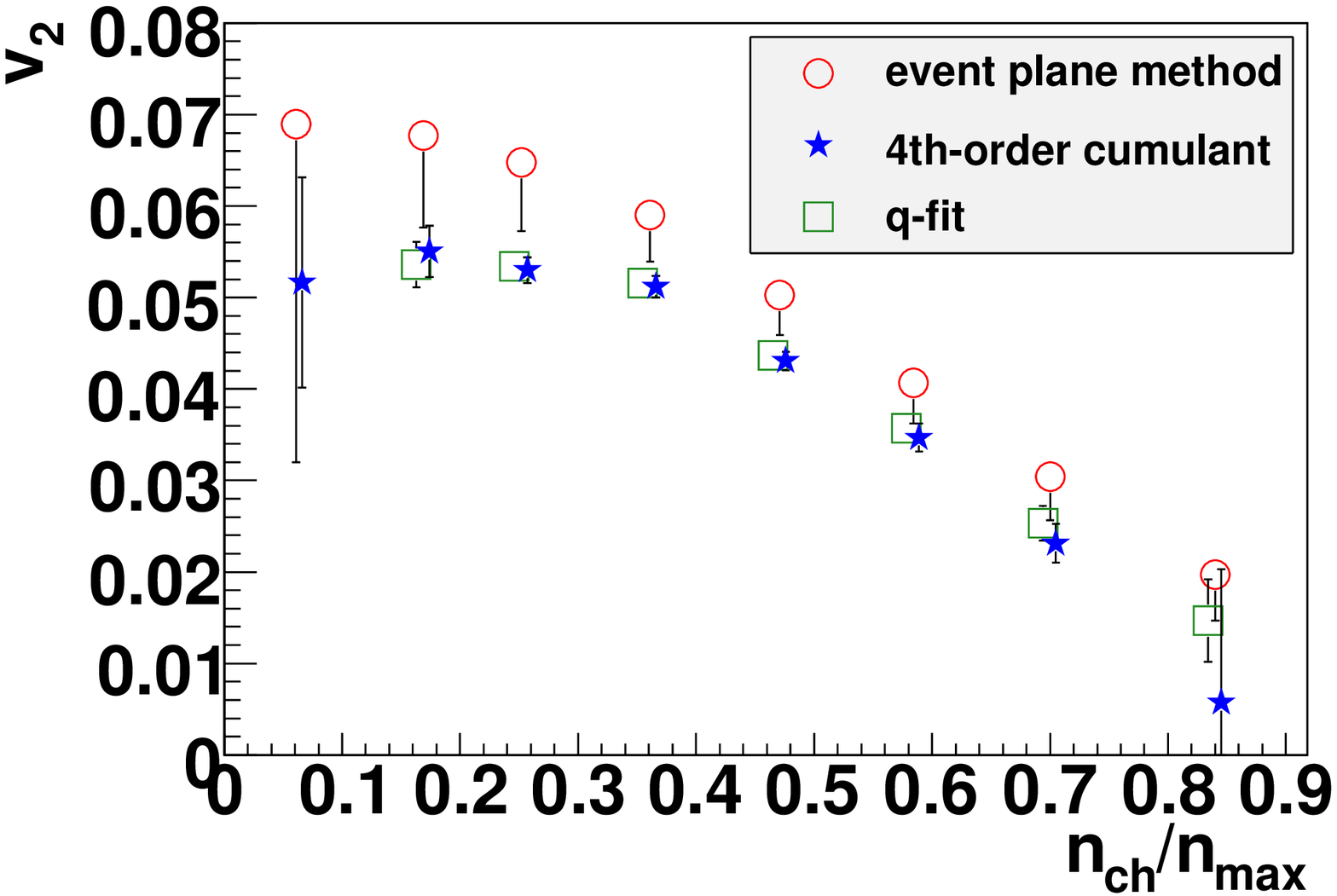}}}
\vspace{-0.5cm}
\caption{Left: $v_2$ measurements as a function of centrality. This plot is from~\cite{WangQM05}. Right: $v_2$ of Au+Au collisions at $\sNN=130 \mathrm{GeV}$. This plot is made based on datapoints from ~\cite{130GeVPaper} \label{fig:v2compare}}
\end{center}
\vspace{-0.5cm}
\end{figure}

When both $\eps_x$ and $\eps_y$ has a Gaussian distribution, to the first order this is true and it is supported by Glauber Monte Carlo simulations, then the probability density function for $\epspp$ is 
 \be
  \frac{dn}{d\epsp}=\frac{\epsp}{\sgmeps^2} 
  I_0 \left( \frac{\epsp \mean{\epsrp} }{\sgmeps^2} \right) 
  \exp\left(-\frac{\epsp^2 +\mean{\epsrp}^2}{2\sgmeps^2}\right)
  \equiv \BG(\epsp;\mean{\epsrp},\sgmeps),
\label{eqbg}
\ee
With this p.d.f., one can show that $\eps_{part}\{4\}=\mean{\eps_{RP}}$~\cite{PlanePaper}. Similarly, under the assumption that $v_2$ is proportional to the initial 
system eccentricity (this is not true over a broad centrality range, but for a fine centrality bin, it still ensures a good Gaussian for $v_2$ and $s_2(\equiv\mean{\mathrm{sin2}(\phi-\Psi_{RP})})$, which is what is required for the derivation of BG formula shown in Eq.(~\ref{eqbg}) ), one can find that $\vfour=\mean{v_2\{RP\}}$~\cite{PlanePaper,EccPaper}. That means that, $\vfour$ and $\eps_{part}\{4\}$ are measurements sensitive to the reaction plane, not the participant plane. Indeed, that explains the reason that STAR's $v_2\mathrm{\{ZDCSMD\}}$ agrees with $\vfour$, as shown in the left panel of Fig.\ref{fig:v2compare}. $v_2\mathrm{\{ZDCSMD\}}$ is measured with the first order event plane reconstructed by spectator neutrons thus is more sensitive to $v_2$ in the reaction plane (not participant plane).
%--------------------------------------------------------------------------
%--------------------------------------------------------------------------
Because the p.d.f. of the magnitude of the flow vector, the
$q-$distribution~\cite{130GeVPaper}, shares an almost identical formula as Eq.~(\ref{eqbg}), the $v_2$ obtained from fitting the $q-$distribution should be equivalent to $\vfour$, as confirmed by experimental data in the right panel of Fig.\ref{fig:v2compare}. 

To summarize this section, we find that $\vfour$, $v_2\mathrm{\{ZDCSMD\}}$ and $v_2\mathrm{\{q-dist\}}$ are sensitive to anisotropy in the reaction plane, they should be scaled with the standard eccentricity or 4-particle cumulant eccentricity. Other $v_2$ measurements that are based on two particle correlations, that includes $v_2\{2\}$, $v_2\mathrm{\{EP\}}$ and $v_2\mathrm{\{scalarProduct\}}$~\cite{130GeVPaper}, etc., should be scaled with participant eccentricity or the 2-particle cumulant eccentricity.

\section{Hints of Incomplete Thermalization}\label{details}

\begin{figure}[ht]
\begin{center}
\resizebox{
\textwidth}{!}{
\resizebox*{10cm}{6.6cm}{
\includegraphics{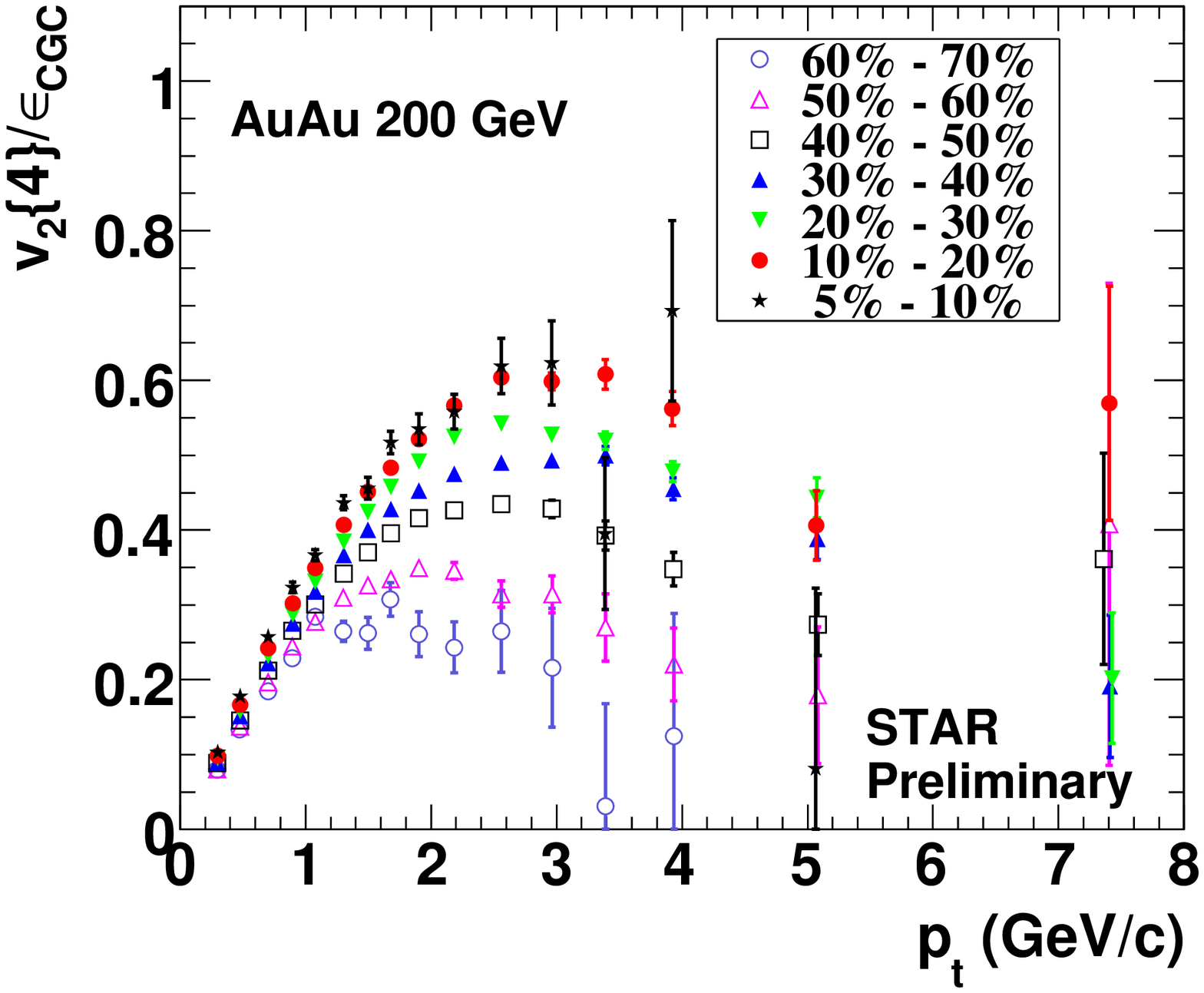}}
\resizebox*{10cm}{6.6cm}{
\includegraphics{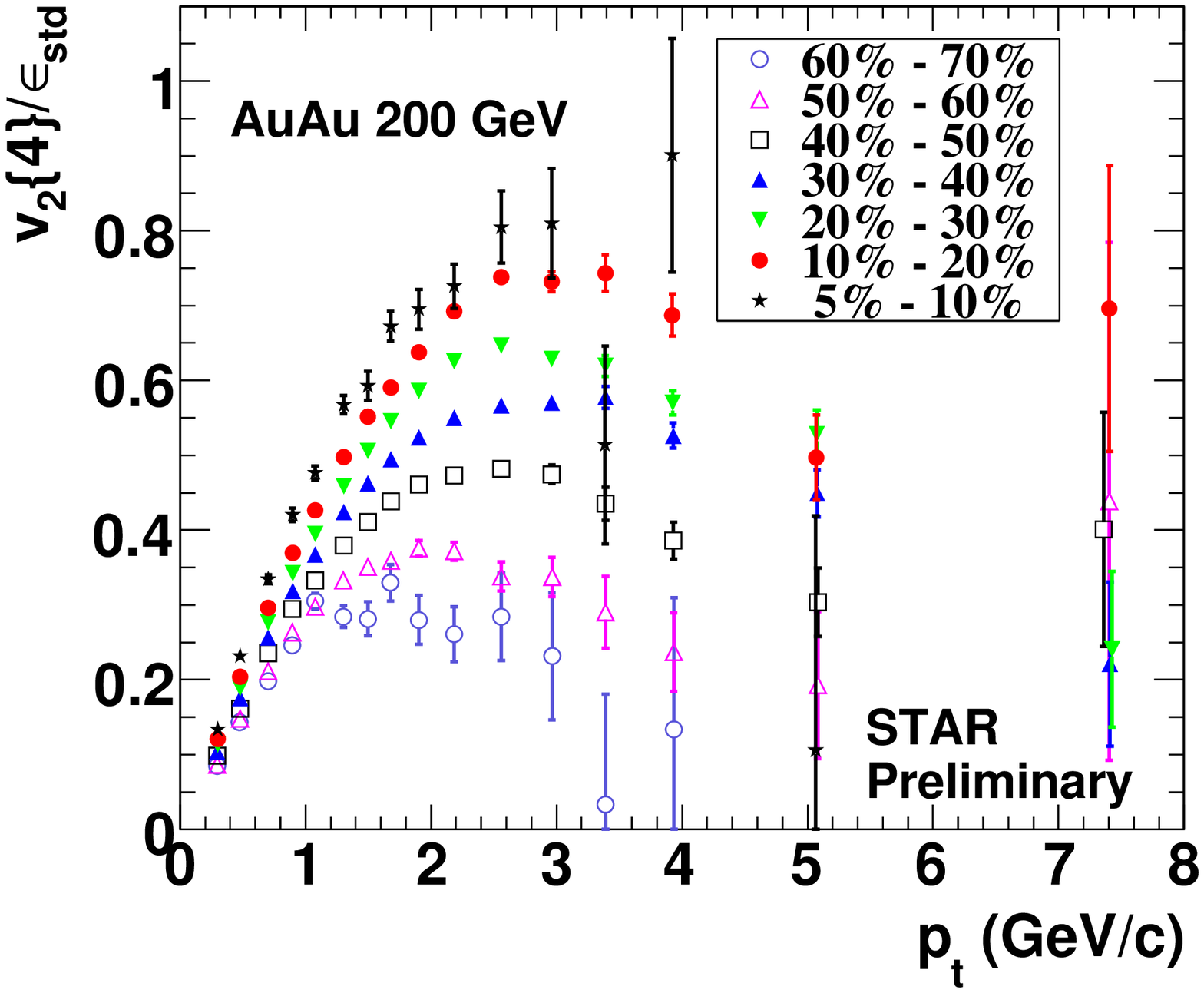}}}
\caption{$v_2$ scaled by initial CGC eccentricity (left) and Glauber eccentricity (right) as a function of $p_t$. This plot is from ~\cite{YutingThesis}. \label{fig:v2pt}}
\end{center}
\vspace{-0.5cm}
\end{figure}

The large data set from run IV Au+Au collisions at $\sNN=200 \mathrm{GeV}$ allows us to extend $\vfour$ measurement to large $p_t$ and in fine centrality bins (see Fig.~\ref{fig:v2pt}). For the reason stated in the previous section, we scale $\vfour$ by initial standard eccentricities. The left plot of Fig.~\ref{fig:v2pt} shows $\vfour$ scaled by the eccentricity from Color Glass Condensate (CGC)~\cite{Drescher}, and the right plot shows $\vfour$ scaled by the Monte Carlo Glauber eccentricity for wounded nucleons. As expected, for the CGC case, the magnitude of $v_2/\epsilon$ is lower if compared to the ratio in which a Glauber eccentricity is used. For both cases, we see the ratio rises from peripheral events to central events, indicating that stronger flow has been developed in central collisions. We also notice that the $p_t$ where $v_2$ reaches its maximum increases from peripheral collisions to central collisions, which is consistent with the expectation that the applicable range for hydrodynamics extends to large pt in central collisions. Note that $v_2/\epsilon$ shows sign of saturation in central collisions for the CGC case but not much for the Glauber case. This is explained by~\cite{Drescher} as the following: in very peripheral collisions, due to little asymmetry in the saturation scales, the CGC eccentricity approaches the same value as that in the Glauber model, but in central collisions, CGC predicts a larger eccentricity than the Glauber model when there is a large asymmetry in the local saturation scales of the collisions partners, along a path in impact-parameter direction away from the origin. Note that $v_2\{2\}$ is not suitable for this study because it is more susceptible to nonflow at large $p_t$.
%--------------------------------------------------------------------------
%--------------------------------------------------------------------------

To understand how well hydrodynamics describes STAR's $v_2$, we investigated the behavior of $v_2$ under the contex of the transport model, which will be reduced to hydrodynamics when the mean free path is much smaller than the system size~\cite{OllitraultBoltzmannEquation}. In such approach, the dependence of $v_2/\epsilon$ on particle's density in the transverse plane ($1/S dN/dy$) can be described by:
\be
  \frac{v_{2}}{\epsilon}= \left[ \frac{v_{2}}{\epsilon} \right]_{hydro} \frac{1}{1+K/K_0}= \left[ \frac{v_{2}}{\epsilon} \right]_{hydro} \frac{1}{1+\left( \sigma \frac{c_s}{c} \frac{1}{4S}\frac{dN}{dy} \right )^{-1} \frac{1}{K_0}}
\label{eqv2density}
\ee
Where $K$ is Knudsen number defined by the mean free path divided by the system size(sometimes it is more convenient to use $K^{-1}$ which means number of collisions a particle encounters before it escapes), and $K_0$ is a constant can be determined through transport calculations. In our study, we take $K_0=0.7$ following the suggestion of Ollitrault~\cite{Drescher}. A factor of 4 in front of $S$ is to take into account the different definition of $S$ between STAR ($S=\pi \sqrt{\mean{x^2}\mean{y^2}}$) and~\cite{Drescher}. (($S=4 \pi \sqrt{\mean{x^2}\mean{y^2}}$). $\left[ \frac{v_{2}}{\epsilon} \right]_{hydro}$ and $\sigma$ are free parameters that have to be determined from fitting the data. In this approach hydrodynamic limit of $v_2/\epsilon$ can be never reached, but can be only asymptotic approached. 

%-----------------------------------------------------------------------
\begin{figure}
\vspace{-0.2cm}
  \begin{center}
    \begin{minipage}[t]{0.48\linewidth}
\resizebox*{7cm}{4.5cm}{
\includegraphics[]{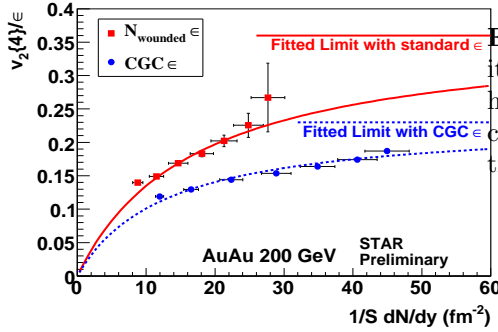}}
    \end{minipage}\hfill
    \begin{minipage}[t]{0.48\linewidth}
    \vspace{-4.4cm}  \caption{$v_2\{4\}$ scaled by initial eccentricities, as a function of centrality. Fitted hydrodynamic limits for  $v_2/\epsilon$ are indicated by horizontal lines. CGC eccentricity and overlap area S are from~\cite{Drescher}. \label{fig:v2density}}
    \end{minipage}
  \end{center}
\vspace{-0.8cm}
\end{figure}
%-----------------------------------------------------------------------

Fig.~\ref{fig:v2density} shows $\vfour$ scaled by CGC initial eccentricity and Glauber initial eccentricity for wounded nucleons, as a function of particle density in the transverse plane, with fits to Eq.~\ref{eqv2density}. The fitted hydrodynamic limits for the ratio of $v_2/\epsilon$ are $0.23$ and $0.36$, for CGC case and Glauber case, respectively. For the same reason mentioned above, we see that the curve shows a hint of saturation for the CGC case, but not for the Glauber case, due to the relatively larger $\epsilon$ for CGC case in central collisions. It is interesting to see that, for central Au+Au collisions, the ratio of $v_2/\epsilon$ is about $20-30\%$ away from hydrodynamic limits. It means that, there is still significant room for flow to grow before the system saturates at hydrodynamic limits.

\begin{figure}
\vspace{-0.2cm}
  \begin{center}
    \begin{minipage}[t]{0.48\linewidth}
\resizebox*{7cm}{4.5cm}{
\includegraphics[]{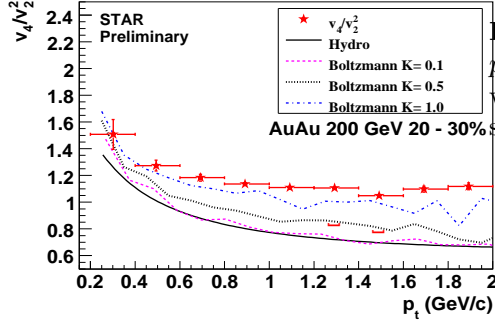}}
    \end{minipage}\hfill
    \begin{minipage}[t]{0.48\linewidth}
    \vspace{-4.4cm}  \caption{$v_4\mathrm{\{EP_2\}}/v_2^2\{4\}$ as a function of $p_t$. This plot is from ~\cite{YutingThesis} with an advanced estimation on systematics represented by half box-brackets. \label{fig:v4v22}}
    \end{minipage}
  \end{center}
\vspace{-1.cm}
\end{figure}

From the simple observation that both $v_2$ and $v_4$ are proportional to $K^{-1}$ for small $K^{-1}$, one expects that $v_4/v_2^2$ decreases with $K^{-1}$, reaching a minimum when the hydrodynamical regime is reached. For this reason the ratio of $v_4/v_2^2$ has been argued as a probe to test the degree of thermalization. Fig.~\ref{fig:v4v22} shows STAR's measurement of $v_4/v_2^2$ as a function of transverse momentum. The major systematic uncertainty in this measurement comes from $v_2$~\cite{YutingThesis}. In this analysis, the induced systematic error from $v_2$ uncertainty is estimated by studying the difference between $\vfour$ and $v_2$ measured with event plane constructed by tracks from STAR's Forward Time Projection Chamber (FTPC). This advanced study reduces the previously-reported systematic error at QM06 conference by $40\%$ relatively. The dashed lines are ratio come out of calculations by solving Boltzmann equations with Monte Carlo simulation, with different Knudsen number $K$. When the Knudsen number is small, it recovers the hydrodynamic limit as indicated the solid line. The plot shows that the system exhibits, again, significant deviation from ideal hydrodynamic limit ($K<<1$), and the data is consistent with a incomplete thermalized system with $K>0.5$.
%-----------------------------------------------------------------------
%-----------------------------------------------------------------------

\section{Conclusion}\label{details}

To summarize, we found that $\vfour$, $v_2\mathrm{\{ZDCSMD\}}$ and $v_2\mathrm{\{q-dist\}}$ are all sensitive to azimuthal correlation w.r.t the reaction plane, not the participant plane,thus they should be scaled by eccentricities that are sensitive to reaction plane too. That includes standard eccentricity and 4-particle cumulant eccentricity. For $v_2$ methods that are based on two particle correlations, they are sensitive to the azimuthal correlation w.r.t the participant plane, and they need to be scaled by the corresponding eccentricities that are sensitive to the participant plane. That includes participant eccentricity or 2-particle cumulant eccentricity. We found that from peripheral to central Au+Au collisions flow increases, and the applicable range for hydrodynamics extends to larger $p_t$. However, $v_2/\epsilon$ and $v_4/v_2^2$ shows significant deviation from ideal hydrodynamic limit, when that limit is extracted from fitting the data itself with a Boltzmann equation motivated formula. Our study shows that although in general hydrodynamic does a good job in terms of describing $v_2$ at RHIC, there are features that are not consistent with a complete thermalization and they cannot be easily dismissed.

\vfill\eject
\end{document}